\documentclass{article} %
\usepackage{betterbodies,times}

\usepackage{hyperref}
\usepackage{url}

\title{BetterBodies: Reinforcement Learning guided Diffusion for Antibody Sequence Design}
\iclrfinalcopy

\author{Yannick Vogt$^{1,2}$\thanks{Corresponding author: vogty@cs.uni-freiburg.de}
\And Mehdi Naouar$^{1,2}$ \And Maria Kalweit$^{1,2}$ \AND Christoph Cornelius Miething$^{2,3}$ \And Justus Duyster$^{3}$ \AND Joschka Boedecker$^{1,2,4}$ \And Gabriel Kalweit$^{1,2}$ \AND
\\
$^1$University of Freiburg, $^2$Collaborative Research Institute Intelligent Oncology (CRIION)\\ $^3$University Medical Center Freiburg, $^4$BrainLinks-BrainTools
}

\usepackage{soul}
\usepackage{multirow}
\newcommand{\basic}{basic}
\newcommand{\our}{our method}
\newcommand{\Our}{Our method}
\newcommand{\ours}{our methods}

\newcommand{\random}{random}
\newcommand{\murine}{natural}
\newcommand{\expert}{expert}

\newcommand{\filtering}{filtering}
\newcommand{\contrastive}{contrastive latent}

\newcommand{\absolut}{Absolut!\ }

\newcommand{\tabref}[1]{Table~\ref{#1}}
\usepackage{glossaries}
\newacronym{rl}{RL}{Reinforcement Learning}
\newacronym{bc}{BC}{behavior cloning}
\newacronym{elbo}{ELBO}{evidence lower bound}
\newacronym{vae}{VAE}{Variational Autoencoder}
\newacronym[plural=AMPs,firstplural=antimicrobial peptides (AMPs)]{amp}{AMP}{antimicrobial peptide}
\newacronym{ai}{AI}{artificial intelligence}
\newacronym[plural=GFlowNets, firstplural=Generative Flow Networks (GFlowNets)]{gflow}{GFlowNet}{Generative Flow Network}
\newacronym[plural=MDPs, firstplural=Markov Decision Processes (MDPs)]{mdp}{MDP}{Markov Decision Process}
\newacronym[plural=CDRs, firstplural=Complementarity Determining Regions (CDRs)]{cdr}{CDR}{Complementarity Determining Region}
\newacronym{sql}{SQL}{Structured Q-learning}
\newacronym{rr}{RR}{Replay Ratio}
\newacronym{auc}{AUC}{Area under the Curve}
\newacronym[plural=AAs,firstplural=amino acids (AAs)]{aa}{AA}{amino acid}
\newacronym{dqn}{DQN}{Deep Q-networks}
\usepackage{graphicx}
\usepackage{amssymb}
\usepackage{adjustbox}
\usepackage{multirow}

\begin{document}

\maketitle

\begin{abstract}
Antibodies offer great potential for the treatment of various diseases. However, the discovery of therapeutic antibodies through traditional wet lab methods is expensive and time-consuming. The use of generative models in designing antibodies therefore holds great promise, as it can reduce the time and resources required. Recently, the class of diffusion models has gained considerable traction for their ability to synthesize diverse and high-quality samples. In their basic form, however, they lack mechanisms to optimize for specific properties, such as binding affinity to an antigen. In contrast, the class of offline \gls{rl} methods has demonstrated strong performance in navigating large search spaces, including scenarios where frequent real-world interaction, such as interaction with a wet lab, is impractical.  Our novel method, BetterBodies, which combines Variational Autoencoders (VAEs) with \gls{rl} guided latent diffusion, is able to generate novel sets of antibody CDRH3 sequences from different data distributions. Using the \absolut simulator, we demonstrate the improved affinity of our novel sequences to the SARS-CoV spike receptor-binding domain. 
Furthermore, we reflect biophysical properties in the VAE latent space using a contrastive loss and add a novel Q-function based \filtering{} to enhance the affinity of generated sequences. 
In conclusion, methods such as ours have the potential to have great implications for real-world biological sequence design, where the generation of novel high-affinity binders is a cost-intensive endeavor.
\end{abstract}

\section{Introduction}
Antibodies are a class of proteins with great potential for treating diseases such as cancer~\citep{kaplon2023antibodies,norman2019computational, robert_unconstrained_2022}. However, the discovery of therapeutic antibodies in classical wet lab experiments is constrained by high costs and low throughput~\citep{angerm_uller2020model, angermueller2020,shanehsazzadeh_unlocking_2023}. Computational antibody design using generative models, therefore, holds immense potential for reducing the time and resources needed~\citep{shanehsazzadeh_unlocking_2023}. 

Diffusion models have recently received considerable attention due to their ability to generate diverse and high-quality data~\citep{pml2Book}. Their versatility makes them applicable to numerous tasks in the realm of protein design, including protein structure prediction~\citep{anand_protein_2022}, protein-protein docking~\citep{ketata_diffdock-pp_2023}, and protein sequence design~\citep{chen_amp-diffusion_nodate}. However, \basic{} diffusion is not capable of optimizing for a desired property such as binding affinity to an antigen. In contrast, \gls{rl} methods have demonstrated remarkable efficacy in identifying solutions in large search spaces~\citep{silver_mastering_2016}. In the domain of offline \gls{rl}, the objective is to learn an optimal policy from a pre-collected dataset without any real-world interaction. This is well-suited to antibody sequence design, where direct access to a wet lab is not feasible.
Consequently, the combination of diffusion models and (offline) \gls{rl} methods has great potential for the field of computational antibody design.
In their work,~\citet{wang_diffusion_2022} demonstrated that \gls{rl} methods can be utilized to guide continuous diffusion models toward optimal regions within the explored space.

To extend recent advances in offline \gls{rl} to the field of antibody design, we frame the antibody sequence design task as a stepwise \gls{aa} placement task. This stepwise approach facilitates the stitching of parts of sub-optimal sequences to create improved sequences~\citep{kumar2022should}.
The \glspl{aa} are mapped into a continuous latent space using a \gls{vae}. The \gls{rl} policy, represented by a continuous diffusion model guided by a learned Q-function, is then trained to generate the optimal next \gls{aa} conditioned on the previously placed \glspl{aa}.

Our contributions are as follows: We propose BetterBodies, a novel method for antibody CDRH3 sequence design, which given a set of training sequences, is able to generate diverse sequences with improved binding affinity in the Absolut! benchmark~\citep{robert_unconstrained_2022}. In experiments, we demonstrate that our method can learn from a variety of distributions of sequences and affinity values, including random sequences, sequences generated by an \gls{rl} agent, and murine antibody sequences.
Additionally, we propose a novel \filtering{} method, using the learned Q-function as a discriminator, as well as a latent space regularization to represent biophysical properties in the VAE latent space. Both methods can be shown to further enhance the average affinity of returned sequences.

\section{Background}
\label{sec:background}

In this section, we provide the necessary background on latent diffusion models, our \gls{mdp} formulation of antibody design, \gls{rl}, and \glspl{vae}.

\subsection{Antibody Sequence Design}
Antibodies are a class of proteins, consisting of a sequence of \glspl{aa}, utilized by the immune system to recognize and bind foreign molecules (antigens) with high specificity~\citep{norman2019computational, robert_unconstrained_2022}. Due to their favorable binding properties, they have become the leading class of new drugs developed~\citep{lu2020development, norman2019computational}. 

Given that there are 20 possible \glspl{aa} to be placed at each sequence position, the total search space for sequences of length $L$ contains $20^L$ sequences. However, it has been demonstrated that specific regions of the antibodies, the so-called \glspl{cdr}, contain the majority of antigen-binding \glspl{aa}~\citep{norman2019computational}. Furthermore, the third \gls{cdr} of the heavy chain (CDRH3) has been shown to have the largest influence on the antibodies' specificity~\citep{xu2000diversity}. Consequently, we utilize the design of CDRH3 sequences as a proxy for the design of complete antibodies. The \absolut software, which we employ to approximate antibody CDRH3 binding affinity to an antigen, fixes the length of this region to 11 positions. Thus, in this work, we will set the length of designed sequences to $L=11$, resulting in approximately 205 trillion possible sequences. This vast space precludes exhaustive search, thereby underscoring the potential impact of computational antibody design on wet labs.

\subsection{Variational Autoencoders}
Autoencoders are encoder-decoder networks trained to minimize a reconstruction loss between their input $x$ and reconstructed input $d_{\rho}(e_{\omega}(x))$, where $e_\omega$ is the encoder network and $d_\rho$ is the decoder network represented by their learnable parameters $\omega$ and $\rho$.
The \gls{vae}~\citep{kingma_auto-encoding_2022} is a specific type of autoencoder in which the continuous latent representation, denoted by $z=e_\omega(x)$, follows a probabilistic distribution $p_\omega(z|x)$. 
The latent representation $z\sim \mathcal{N}(\mu_\omega^x, \sigma_\omega^x)$ is typically defined as a Gaussian distribution with a learned mean $\mu_\omega$ and standard deviation $\sigma_\omega$.
In addition to the reconstruction loss, the \gls{vae} is regularized such that the latent distribution minimizes the Kullback-Leibler (KL) divergence to a Gaussian distribution $\mathcal{N}(0, \mathbf{I})$, facilitating a dense latent space. 
In our setting, we use \glspl{vae} to encode \glspl{aa} classes into a two-dimension latent space and use a Binary Cross Entropy loss for reconstruction.
Furthermore, \glspl{vae} can be regularized to cluster inputs belonging to the same class by pulling them together in embedding space, while simultaneously pushing apart clusters of inputs from different classes~\citep{khosla_supervised_nodate}.

\subsection{Diffusion Models}

Diffusion models employ a \textit{forward process}, or \textit{diffusion process}, to gradually corrupt observed data into noisy data and learn a \textit{reverse process}, or \textit{denoising process}, to undo the corruption. %
A trained model can thus be used to generate high-quality data from noise~\citep{pml2Book}.

In this work, we are dealing with both diffusion steps $n\in\{0,..,N\}$ and time steps $t\in\{0,...,T\}$. To facilitate clarity, we will use superscripts for diffusion steps and subscripts for time steps. 
Diffusion probabilistic models~\citep{ho_denoising_nodate, pmlr-v37-sohl-dickstein15} are a class of latent variable models defined as $p_\theta(x^0) := \int p_\theta(x^{0:N} )dx^{1:N}$. Here, $x^1, ..., x^N$ are latent variables of the same dimensionality as the data sample $x^0$ drawn from the observed data distribution $q(x^0)$. In our setting, these data samples are two-dimensional embeddings of \glspl{aa} drawn from a \gls{vae} latent space. The forward process gradually adds Gaussian noise to $x^0$ according to a noise schedule $\beta^1, ..., \beta^N$, over $N$ steps~\citep{ho_denoising_nodate}. In particular, the forward process is defined as $
    q(x^{1:N} |x^0) := \prod^N_{n=1} q(x^n|x^{n-1})
$, with single step transition
$
     q(x^n|x^{n-1}) := \mathcal{N}(x^n; \sqrt{1 - \beta^n}x^{n-1}, \beta^n \mathbf{I})
$.

The reverse process is the joint distribution $p_\theta(x^{0:N})$ defined as a Markov chain starting at $p(x^N) = \mathcal{N}(x^N ; 0, \mathbf{I})$ given as 
$
    p_\theta(x^{0:N} ) := p(x^N) \prod^N_{n=1} p_\theta(x^{n-1}|x_n),
    \label{eq:reverse}
$
with a learned Gaussian transition $p_\theta(x^{n-1}|x^n)$.
The objective of training $p_\theta$ is to maximize the expected log-likelihood of the data, given by the \gls{elbo} $\mathbb{E}_q[\log \frac{p_\theta(x^{0:N})}{q(x^{1:N} |x^0) } ]$. 
In essence, the objective is to maximize the probability of reconstructing a sample $x^0$ from a noisy sample $x^N$.
In practice, instead of predicting $x^{n-1}$ given $x^n$, a noise prediction model $\epsilon_\theta$ is trained~\citep{ho_denoising_nodate, pml2Book}. Consequently, the loss for the diffusion model given a dataset $D$ can be simplified to 
\begin{equation}
    L(\theta) = \mathbb{E}_{n\sim \text{Unif}(1,N),\epsilon\sim \mathcal{N}(0,\mathbf{I}),x^0\sim D} [||\epsilon - \epsilon_\theta(\sqrt{\Bar{\alpha}^n}x^0 + \sqrt{1-\Bar{\alpha}^n}\epsilon, n)||^2],
    \label{eq:Lbc}
\end{equation}
where  $\alpha^n:=1-\beta^n$ and $\Bar{\alpha}^n:=\prod^n_{i=1} \alpha^i$.

\subsection{Reinforcement Learning}
\label{sec:rl}
In \gls{rl}, tasks are typically formulated as \glspl{mdp}.
We define a deterministic \gls{mdp} as a tuple $\langle S,S_0,A,P,R \rangle$, where
$S$ is the set of possible states, $S_0$ is the set of initial states $S_0 \in S$, $A$ is the set of possible actions $a$ executable in $s\in S$, $P$ is a deterministic transition function $P(s,a): S \times A \mapsto S$, and $R$ is a deterministic reward function $R(s,a): S \times A \mapsto \mathbb{R}$.

In this work, we address the task of designing discrete \gls{aa} sequences, representing antibody CDRH3 sequences targeting a specific antigen. We choose to frame the task as a stepwise generation process where the \glspl{aa} are placed in the sequence one after the other. To evaluate the binding affinity of designed sequences given a specific antigen, we utilize \absolut~\citep{robert_unconstrained_2022} which sets the length of a \textit{complete} sequence to 11. Thus, we define the set of states $S$ as the set of all possible \gls{aa} sequences up to length 11, including the empty sequence.
We define the set $S_0$ as an empty sequence.
The set $A$ is then defined as the set of 20 natural \glspl{aa}. To facilitate notation, the symbol $a$ is used to refer to the action, the \gls{aa} it represents, and its two-dimensional \gls{vae} latent representation. Consequently, we define $P(s,a)= s\mathbin\Vert a$ as the concatenation of the sequence generated thus far with the next \gls{aa}, extending the sequence by one more \gls{aa}. The reward function $R(s,a)$ is defined corresponding to the predicted free energy using the \absolut software. As sequences of length shorter than 11 can not be evaluated, the reward function is sparse, returning the evaluated free energy \absolut$(s\mathbin\Vert a| \text{antigen})$ for sequences of length 11 and a reward of 0 for all shorter sequences.

 The objective in \gls{rl} is to learn a policy $\pi$ that maximizes the expected sum of rewards. The action-value function $Q$ represents this expected sum starting from a given state $s_t$. We define it as follows:
 \begin{equation}
     Q(s_t,a_t) = \mathbb{E}_{\pi}[R(s_t, a_t) + \sum_{i=1}^{10-t} R(s_{t+i}, a_{t+i})|a_{t+i}\sim \pi(s_{t+i})].
 \end{equation}
 
The policy $\pi$ should thus select the action $a$ that maximizes $Q$ for each state $s$. 
As the search space of CDRH3-sequences is enormously huge, we estimate $\pi$ and $Q$ with function approximations $\pi_\theta$ and $Q_\phi(s,a)$, parameterized by $\theta$ and $\phi$ respectively.

In this work, we focus on the offline \gls{rl} setting, where the agent is trained using a pre-collected dataset, which we consider to be more realistic for the antibody design task, as interactive access to a wet lab is not feasible.
The offline setting presents its own set of challenges~\citep{levine2020offline}, mainly erroneous assignment of high Q-values to state-action pairs outside the provided dataset and a resulting distribution shift in the policy. There are multiple approaches to prevent these issues. \Our{} falls in the class of policy regularization, providing an incentive to remain close to the provided dataset.

\section{Related Work}
In recent years the field of protein sequence design has been approached with a variety of generative models and from a multitude of perspectives. Some~\citep{cowen_rivers2022structured, khan2022antbo, vogt2023stable} approach the task in an online setting, where the policy has continuous access to the evaluation metric and can freely explore the design space to find a high-reward sequence using \gls{rl} or bayesian optimization methods. In other settings,  which are sometimes referred to as active learning settings, the generative policy is trained from pre-collected offline datasets for multiple rounds where generated sequences can be evaluated and might be added to the datasets in between rounds~\citep{angerm_uller2020model, angermueller2020, jain_biological_nodate}. In such settings, ensembles of evolutionary algorithms~\citep{angermueller2020}, \gls{rl} algorithms~\citep{angerm_uller2020model}, and \glspl{gflow}~\citep{jain_biological_nodate} have been employed as generative models. 
Lastly, the task can also be approached from a purely offline perspective, where the generative policy is trained only once on a pre-collected offline dataset and then evaluated~\citep{chen_amp-diffusion_nodate,gruver_protein_2023, jain_biological_nodate}.

We approach the task from a purely offline perspective and will present related work from that domain in more detail here. In some of their experiments,~\citet{jain_biological_nodate} utilize \glspl{gflow} to tackle the design of DNA sequences and protein sequences in the offline setting. They thereby utilize a learned reward model to explore beyond the offline data. In addition to generating samples that optimize a desired property, the networks are also trained to generate samples with high uncertainty according to the learned reward model. The choice of \glspl{gflow}, which are trained to generate samples with likelihoods proportional to their reward fraction in the dataset, intuitively allows for the generation of high-reward samples. In practice, this class of networks is hard to train, due to oversampling of low reward trajectories, and the rewards need to be non-linearly scaled to achieve a good performance~\citep{jain_biological_nodate, pmlr-v202-shen23a}.

In their approach,~\citet{chen_amp-diffusion_nodate} utilize a continuous diffusion model to generate entire \gls{amp} sequences in an ESM-2~\citep{lin_evolutionary-scale_2022} latent space. 
The choice of diffusion models, which are capable of modeling complex multi-modal distributions~\citep{ho_denoising_nodate, wang_diffusion_2022}, appears well suited for the complex circumstances underlying \gls{aa} sequence design.
They demonstrated that generated peptides exhibited similar physicochemical properties to natural peptides and aligned closely with respect to \gls{aa} diversity, which highlights the expressive power of their method. However, they do not employ any technique guiding the diffusion process towards improved sequences.
In contrast,~\citet{gruver_protein_2023} employ discrete diffusion, whereby sequences are directly diffused in the discrete sequence space. They propose guiding the diffusion model by a learned value function. However, their formulation requires the diffusion model and the value function to share some of their hidden layers and requires the value function to be trained on corrupted inputs~\citep{gruver_protein_2023}.
Furthermore, wet lab experiments were conducted on generated antibody sequences, which indicated that some of the designed sequences may have improved real-world binding affinity.

\gls{rl} methods have been demonstrated to identify solutions in large search spaces~\citep{silver_mastering_2016} and to be applicable to the sequence design task~\citep{angerm_uller2020model,cowen_rivers2022structured, vogt2023stable}. Moreover, it has been shown that learned Q-functions, employed in \gls{rl}, can be utilized to guide continuous diffusion models towards high rewards in offline \gls{rl}~\citep{wang_diffusion_2022}.

In our work, we apply recent advances in offline \gls{rl} to the protein sequence design task. Similar to~\citet{jain_biological_nodate} but in contrast to~\citet{gruver_protein_2023} and~\citet{chen_amp-diffusion_nodate} we choose to phrase the protein sequence as a stepwise \gls{aa} generation task, conditioned on the sequence generated thus far, instead of generating entire sequences at once. Such an approach facilitates the use of Q-functions which are able to stitch together improved sequences from suboptimal ones~\citep{kumar2022should}. We enable continuous diffusion of discrete \glspl{aa} by training a \gls{vae} encoding discrete \gls{aa} into a latent space and decoding generated continuous vectors back to the discrete \glspl{aa}. 
Furthermore, we show how biophysical properties can be injected into the latent space to improve performance. Finally, we propose a novel \filtering{} method based on learned Q-values to enhance the average affinity in the set of returned sequences.

\section{BetterBodies}
\label{sec:method}
\begin{figure*}[h!]
\centering
\includegraphics[width=1.0\linewidth]{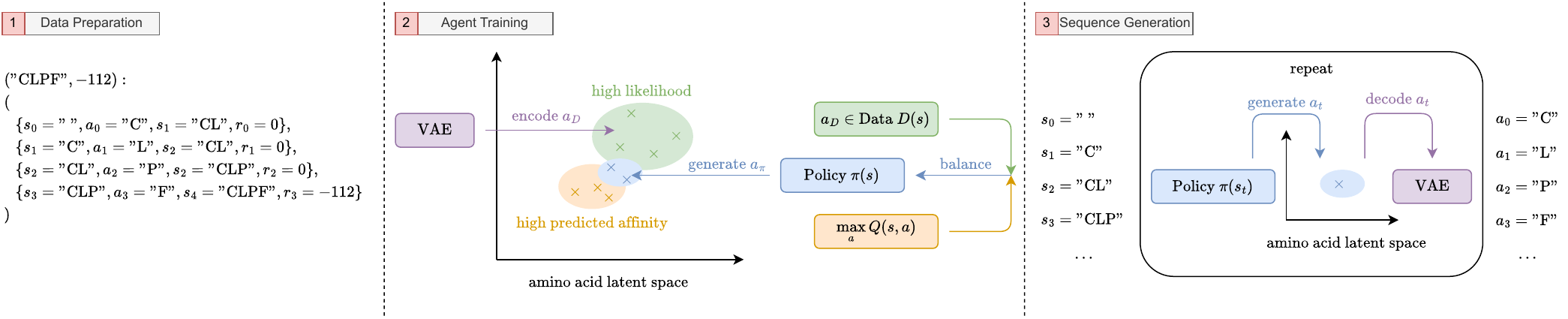} 
\caption{Overview over \our{} on a fictional sequence of length 4. 
(1) A given dataset comprising sequence-affinity pairs is transformed into subsequences ($s$) and actions ($a$) which extend those sequences with additional amino acids, together with rewards representing the affinity of the full sequences.
(2) \Our{} utilizes a \gls{vae} to encode  \glspl{aa} into a two-dimensional latent space. The diffusion policy $\pi$ is trained to generate a latent \gls{aa} vector $a_\pi$ given an incomplete amino acid sequence $s$. We balance the policy between generating \glspl{aa} with high likelihood given the training dataset $D$ and \glspl{aa} that maximize a learned Q-function, which predicts sequence affinity to a given antigen. (3) By repeating the generative process, \glspl{aa} are iteratively concatenated to generate a sequence. In each timestep $t$ the policy $\pi$ generates a latent vector $a_t$ given $s_t$. Subsequently, the \gls{vae} decodes the \gls{aa}, which is then concatenated to $s_t$ to generate $s_{t+1}$.}
\label{fig:summary}
\end{figure*}
The objective of \our{}, BetterBodies, summarized in \Figref{fig:summary}, is to train a policy $\pi$ that, in a stepwise manner, generates high-affinity CDRH3 sequences by concatenating generated \glspl{aa} given an initial set $D$ of sequence-affinity pairs. 
Furthermore, the generated sequences should be novel and diverse. Diffusion models have recently gained popularity due to their ability to model complex distributions and generate diverse and high-quality samples~\citep{ho_denoising_nodate,pml2Book, wang_diffusion_2022}. Consequently, we represent the policy $\pi_\theta$ using a continuous latent diffusion model with parameters $\theta$.

\subsection{Continuous Amino Acid Representations and Encoding Biophysical Properties}
In contrast to discrete diffusion~\citep{gruver_protein_2023}, which is designed to directly generate categorical values in the reverse process, our policy $\pi$ generates a continuous latent vector. This continuous vector facilitates to guide the diffusion model and shape to latent space to incorporate biophysical properties but requires representing the categorical \glspl{aa} as continuous vectors. We choose to represent them using the two-dimensional latent vectors of a \gls{vae} which we train and freeze before training the diffusion model. The latent vectors generated by $\pi$ are then transformed back to discrete \glspl{aa} using the \gls{vae}'s decoder network.
To train the \gls{vae}, each \gls{aa} $a$ is represented as a one-hot vector and mapped to a two-dimensional latent $z=e_\omega(a)\sim \mathcal{N}(\mu_\omega^a, \sigma_\omega^a)$ using the encoder network $e_\omega$. 
The decoder network $d_\rho(z)$ then maps the latent vector $z$ back to a probability distribution over discrete \glspl{aa}. Consequently, the \gls{vae} can be trained end-to-end by minimizing the Binary Cross Entropy (BCE) loss between the input $a$ and the decoder's output. Additionally, the distribution of latent variables $z$ is regularized to minimize the KL divergence to the Gaussian distribution. The loss function of the \gls{vae} is then given as $L(a) = \text{BCE}(a, d_\rho(e_\omega(a))) +\text{KL}(\mathcal{N}(\mu_\omega^a, \sigma_\omega^a), \mathcal{N}(0, \mathbf{I}))$.

Furthermore, the latent space utilized in \our{} also allows for incorporating additional biases. As a proof of concept, we chose to regularize the \glspl{vae} latent space to represent \gls{aa} properties. Specifically, we group the \glspl{aa} according to their side chains' properties, based on classification by \citet{garrett_biochemistry_2010} with some modifications (cp. Appendix \Secref{app:groups}).
In this modification of \our{}, which we refer to as BetterBodies-C(ontrastive), we added a supervised contrastive loss to the \gls{vae} training objective to realize this grouping in latent space~\citep{khosla_supervised_nodate}.
Specifically, the contrastive loss is given by
\begin{equation}
    - \sum_{a \in A} log [\frac{1}{|\text{group(}a\text{)}|} \sum_{p \in \text{group(}a\text{)}} \frac{\exp(z_a \cdot z_{p})/ \tau}{\sum_{a' \in A \setminus a} \exp(z_a \cdot z_a') / \tau}],
\end{equation}
where $A$ is the set of all \glspl{aa}, group($a$) represents the subset of \glspl{aa} belonging to the same group as $a$, $\cdot$ represents the cosine similarity over latent representations $z$, and $\tau$ is a temperature hyperparameter. This loss maximizes the similarity between \glspl{aa} in the same group and maximizes it in between groups.

Recall from \Secref{sec:background}, that the diffusion's reverse process starts with $x^N\sim \mathcal{N}(x^N ; 0, \mathbf{I})$. Consequently, the regularization term $\text{KL}(\mathcal{N}(\mu_\omega^a, \sigma_\omega^a), \mathcal{N}(0, \mathbf{I}))$ also prevents a mismatch between the \gls{aa} latent space and the diffusion model's latent space.

\subsection{Guiding Diffusion Policies using Reinforcement Learning}
The policy $\pi$ is trained to achieve a balance between two objectives: generating latent vectors representing \glspl{aa} with high likelihood given a dataset $D$ and generating \glspl{aa} maximizing a learned Q-function. 
Recall from \Secref{sec:rl}, that we use $a$ to represent an \gls{aa}, its corresponding latent vector, and the corresponding action in the \gls{mdp}.

The loss function corresponding to the first objective, referred to as the \gls{bc} loss, is a slight adaptation of the standard loss function for continuous diffusion models given in \Eqref{eq:Lbc}. In particular, as we generate sequences stepwise, one \gls{aa} $a$ after the other, we condition the diffusion model on the sequence $s$ of \glspl{aa} generated so far. The first loss function thus becomes
\begin{equation}
    L_{BC}(\theta) = \mathbb{E}_{n\sim \text{Unif}(1,N),\epsilon\sim \mathcal{N}(0,\mathbf{I}),(s,a)\sim D} [||\epsilon - \epsilon_\theta(\sqrt{\Bar{\alpha}^n}a + \sqrt{1-\Bar{\alpha}^n}\epsilon, s, n)||^2].
\end{equation}
Simply put, this loss function trains the model to reconstruct the next $a$ given an incomplete sequence $s$ from the dataset $D$. It has been shown that this diffusion approach improves performance on multimodal data in comparison to other training paradigms~\citep{wang_diffusion_2022}.

The \gls{bc} loss alone does not provide a means of generating \glspl{aa} which would result in sequences with improved affinity compared to sequences in $D$.
Consequently, we desire a gradient guiding the policy $\pi_\theta$ towards such \glspl{aa}. 
We follow~\citet{wang_diffusion_2022} and utilize the gradient of a learned Q-function $Q(s,a^0)$ given an incomplete sequence $s$ and an action $a^0$ generated by the policy $\pi_\theta$. The use of a Q-function for guidance in sequence design is promising, as these have been shown to stitch together improved sequences from suboptimal ones and excel in states where taking a specific action is required~\citep{kumar2022should}. 
The full loss for $\pi_\theta$, represented by its learnable parameters $\theta$ is then given as
\begin{equation}
    L(\theta) = L_{BC}(\theta) - \eta \cdot \mathbb{E}_{s\sim D, a^0\sim\pi_\theta}[Q_\phi(s, a^0)].
    \label{eq:balance}
\end{equation}
As $a^0$ is generated using the reverse process of the diffusion model $\pi_\theta$, the gradient of $Q_\phi(s, a^0)$ is propagated through the diffusion model's reverse process, thereby guiding the selection of actions with a high Q-value given the current state $s$. The hyperparameter $\eta$ is used to select a balance between the \gls{bc} loss and maximizing Q-values.
This relatively straightforward combination of a loss function that regularizes the policy towards the dataset and a loss function that facilitates policy improvement beyond the dataset has been demonstrated to be effective in many offline \gls{rl} domains~\citep{fujimoto_minimalist_2021}.

The Q-function $Q_{\phi}$, implemented as clipped double Q-learning~\citep{fujimoto2018addressing}, is trained to minimize the so-called TD-error:
\begin{equation}
                \mathbb{E}_{(s_t ,a_t ,s_{t+1} )\sim D,a^0_{t+1}\sim \pi_\theta'} [|| (R(s_t, a_t) + \min_{i=1,2} Q_{\phi'_i} (s_{t+1}, a^0_{t+1})) 
         - Q_{\phi_i} (s_t, a_t)||^ 2 ],
\end{equation}
where subscripts $t$ indicate the trajectory index (\gls{aa} position).
In practice, the diffusion policy $\pi_\theta$ and the Q-function $Q_\phi$ are updated in an alternating fashion.
\subsection{Filtering generated Sequences}
Finally, we propose a \filtering{} method to enhance the average affinity of returned sequences, referred to as BetterBodies-F(iltering). Consequently, we refer to \our{} with both \filtering{} and \contrastive{} as BetterBodies-CF.
As stated in \Secref{sec:rl}, we only assign a reward corresponding to the sequence's free energy to full sequences of length 11. Consequently, the Q-value $Q_\phi(s_{10}, a_{10})$ of a sequence of length 10 $s_{10}$ concatenated with the last amino acid $a_{10}$ is trained to predict the free energy. We propose to utilize the learned Q-values as a discriminator and sort generated sequences according to their predicted free energy. This allows for the discarding of high-energy sequences above a given percentile.
If the learned Q-values do correlate with the true affinity (inverse of free energy), this method will be effective in retaining high-affinity sequences.

\section{Experiment Setup}
In the following section, we compare \our{} to \glspl{gflow}~\citep{jain_biological_nodate} on three different data distributions. Note, that the utilized datasets and our source code is included in the supplementary material. Further, results for a second antigen, and implementation details are included in the appendix, \Secref{app:3raj} and \Secref{app:implementation}.
\paragraph{Evaluation Metrics}
Our objective is to train a policy $\pi$ which generates a set of unique novel sequences, denoted $D_{gen}$, given a training dataset $D$. The novel sequences should maximize binding affinity to a given antigen. Affinity can be maximized by minimizing the free energy between the antibody, represented by the generated sequence, and the antigen. Therefore, we want to minimize the free energy computed using the \absolut software. 
Furthermore, generated sequences should be diverse and novel in comparison to the dataset $D$. We utilize the definition of diversity and novelty proposed by~\citet{jain_biological_nodate}:
$
    Diversity(D_{gen}) := \frac{\sum_{x_i\in D_{gen}} \sum_{x_j\in D_{gen} \backslash \{x_i\}} d(x_i,x_j)}{|D_{gen}|  (|D_{gen}|-1)}$
and 
    $Novelty(D_{gen}) := \frac{\sum_{x_i\in D_{gen}} \min_{s_j \in D} d(x_i,s_j)}{|D_{gen}|}$, 
where $d(\cdot, \cdot)$ is the Levenshtein distance measuring the amount of difference between two sequences~\citep{leven}.
These measures provide insight into the average number of pointwise mutations in the sequence relative to other sequences in the generated dataset $D_{gen}$ (Diversity) and their closest relative in the original dataset $D$ (Novelty).

\paragraph{Training Datasets}
To assess the efficacy of our method we train it on three different data distributions. These distributions represent CDRH3 sequences and their respective free energies in complex with the SARS-CoV spike receptor-binding domain (PDB ID 2DD8\_S). We selected this antigen, as prior methods on the \absolut benchmark~\citep{cowen_rivers2022structured,khan2022antbo, vogt2023stable} performed comparably weak on this antigen, indicating a higher complexity in identifying effective binders. We present additional evaluations on a second target in the supplementary material. The first distribution comprises a set of 2500 randomly generated sequences, for which the binding affinity to the SARS-CoV spike receptor-binding domain was predicted using the \absolut software. The second set contains 2753 murine CDRH3 sequences, which were categorized as good but not exceptional binders~\citet{robert_unconstrained_2022}.
The final distribution, comprising 2167 sequences, was gathered during the exploration phase of a Q-learning agent, similar to those described by~\citet{vogt2023stable}. Due to the agent's efficacy, this dataset contains sequences that reach affinity levels beyond those found with murine CDRH3 sequences. We refer to the three datasets as \textit{\random{}}, \textit{\murine{}}, and \textit{\expert{}}.
The three datasets thereby represent data distributions that could occur in applications of \our{}. The \random{} data represents an initial lab screening with random CDRH3 sequences, \murine{} a dataset derived from known natural sequences, and \expert{} a dataset as it could occur in an active-learning setting.

\section{Results}
In the following section, we present the results of our experiments. To develop an intuition, we start with an in-depth analysis of the effect of Q-function guidance on diffusion with respect to the maximization of the \gls{elbo} given the dataset $D$, the training stability, as well as the affinity, novelty, and diversity of generated sequences on the \expert{} dataset.
Subsequently, we compare \our{} to \glspl{gflow}~\citep{jain_biological_nodate} on all three datasets.
All experiments are carried out over five seeds.

\subsection{Effect of Guidance}
\Our{} augments the generative process of a diffusion model with Q-value guidance. 
To achieve an improvement in performance over \basic{} diffusion, it is necessary to find an appropriate balance between two objectives: optimizing the \gls{elbo} (represented by $L_{BC}$) given data $D$ and maximizing the Q-value.
As discussed in \Secref{sec:method}, \Eqref{eq:balance}, this balance can be controlled using hyperparameter $\eta$, where $\eta=0$ deactivates the guidance leading to \basic{} diffusion only optimizing the \gls{elbo}.
\begin{figure*}[h]
    
\begin{center}
\begin{minipage}{0.3\textwidth}
\begin{center}
\centerline{\includegraphics[width=1\textwidth]{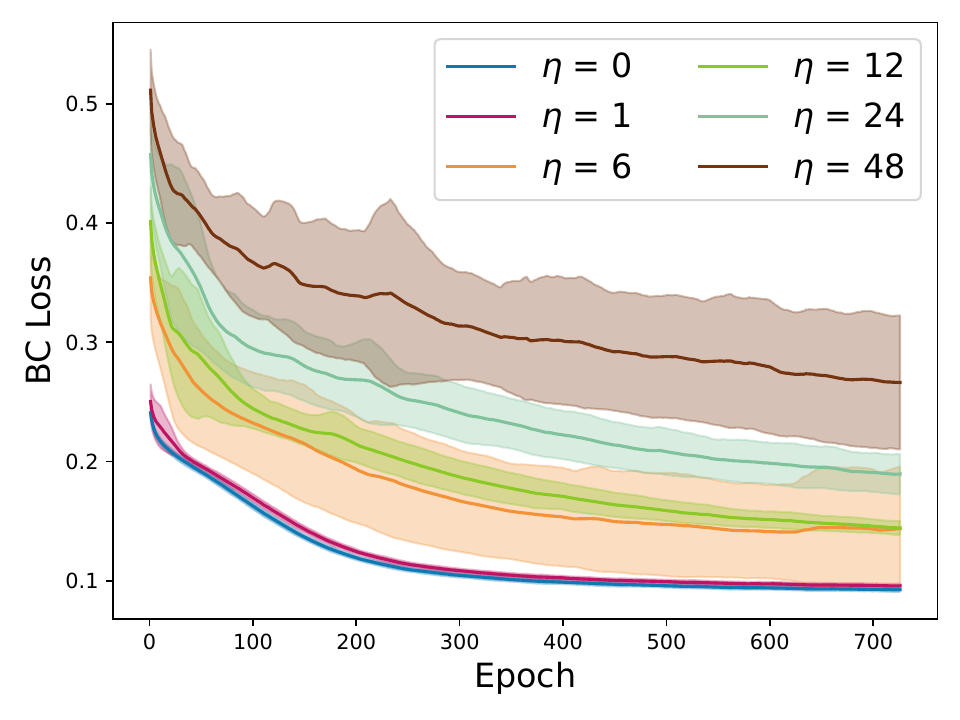}}
\end{center}
\end{minipage}
\begin{minipage}{0.3\textwidth}
\begin{center}
\centerline{\includegraphics[width=1\textwidth]{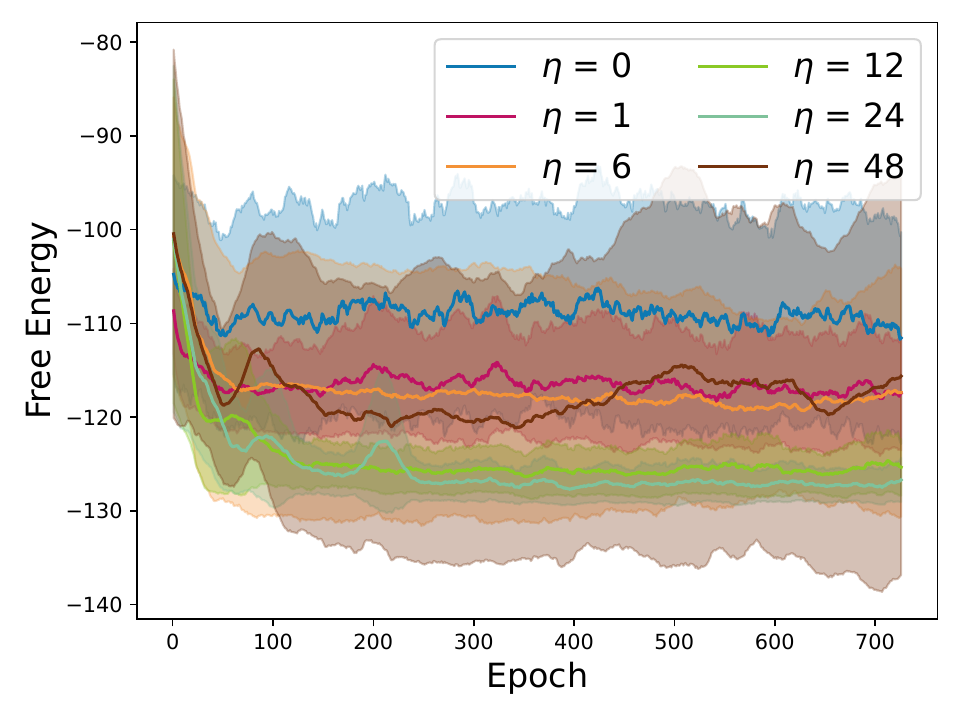}}
\end{center}
\end{minipage}
\\
\begin{minipage}{0.3\textwidth}
\begin{center}
\centerline{\includegraphics[width=1\textwidth]{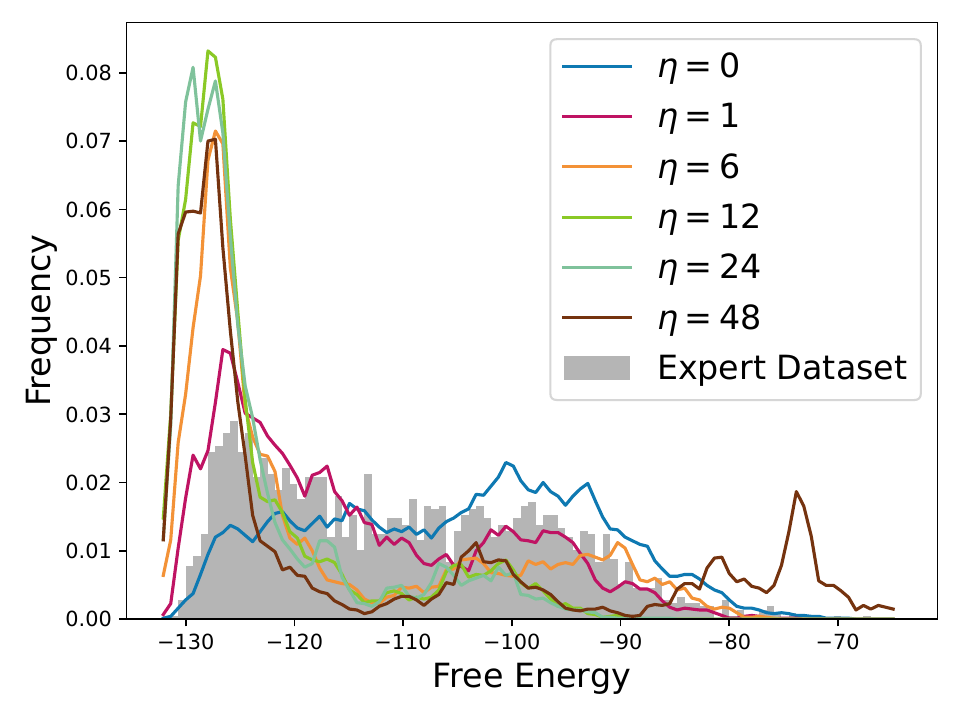}}
\end{center}
\end{minipage}
\begin{minipage}{0.3\textwidth}
\begin{center}
\centerline{\includegraphics[width=1\textwidth]{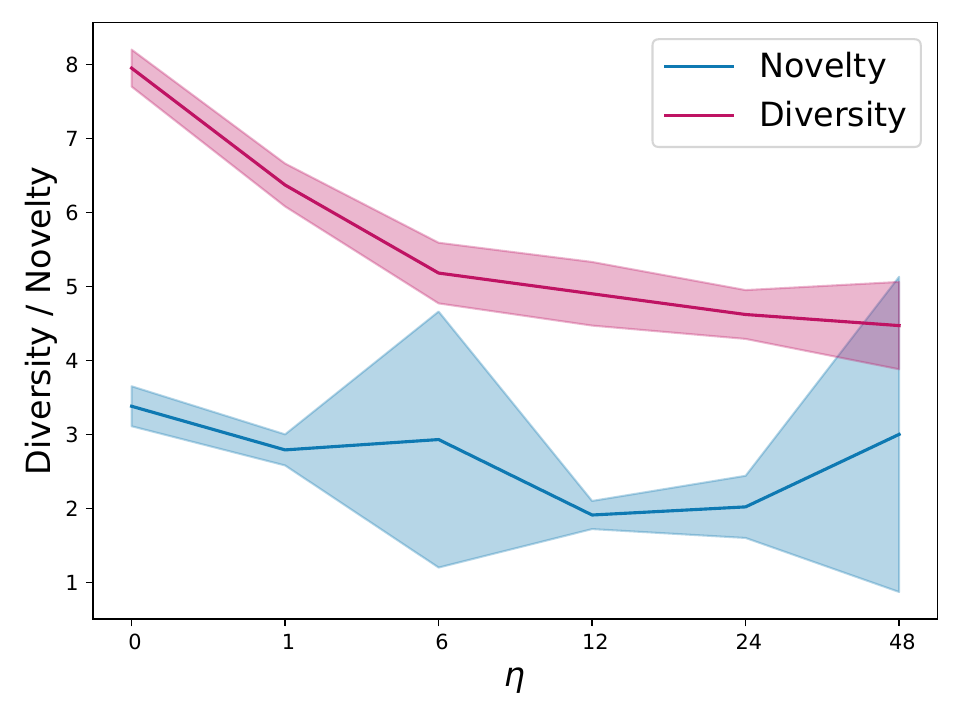}}
\end{center}
\end{minipage}

\end{center}

\caption{The effect of various $\eta$ settings: On the \basic{} diffusion loss $L_{BC}$ (top left), free energy evaluated during training (top right), and free energy distribution of generated unique novel sequences (bottom left), and Diversity and Novelty of generated sequences (bottom right). Distributions of generated sequences are plotted as a running average over three bins.}
\label{fig:ablation_all}
\end{figure*}

In \Figref{fig:ablation_all} (top left), we visualize the effect of varying $\eta$ configurations on the magnitude of $L_{BC}$. 
We observe an increasing trend in $L_{BC}$ when increasing $\eta$, suggesting a shift of the policy away from $D$.

This shift, up to a certain point, corresponds to an increase in the affinity of generated sequences during the training phase, as illustrated in \Figref{fig:ablation_all} (top right). However, with $\eta=48$ training instabilities can be observed.

\Figref{fig:ablation_all} (bottom left), depicts the free energy distribution of unique novel sequences generated after the training phase across multiple $\eta$ settings. While $\eta=0$ roughly matches the training distribution, increasing $\eta$ up to $24$ results in a shift of the distribution towards sequences with low free energy, highlighting the improvement through guidance.

In \Figref{fig:ablation_all} (bottom right) we can visualize the influence of $\eta$ on the diversity and novelty of generated sequences. 
With increasing choice of $\eta$, diversity is decreasing, indicating a guidance towards a narrow distribution of sequences, maximizing the Q-function. For novelty, we can observe a similar trend, which, however, stops with the unstable setting of $\eta=48$ where novelty increases again.

\subsection{Comparison to \glspl{gflow}} %
\begin{figure*}[h]
    
\begin{center}

\includegraphics[width=0.8\textwidth]{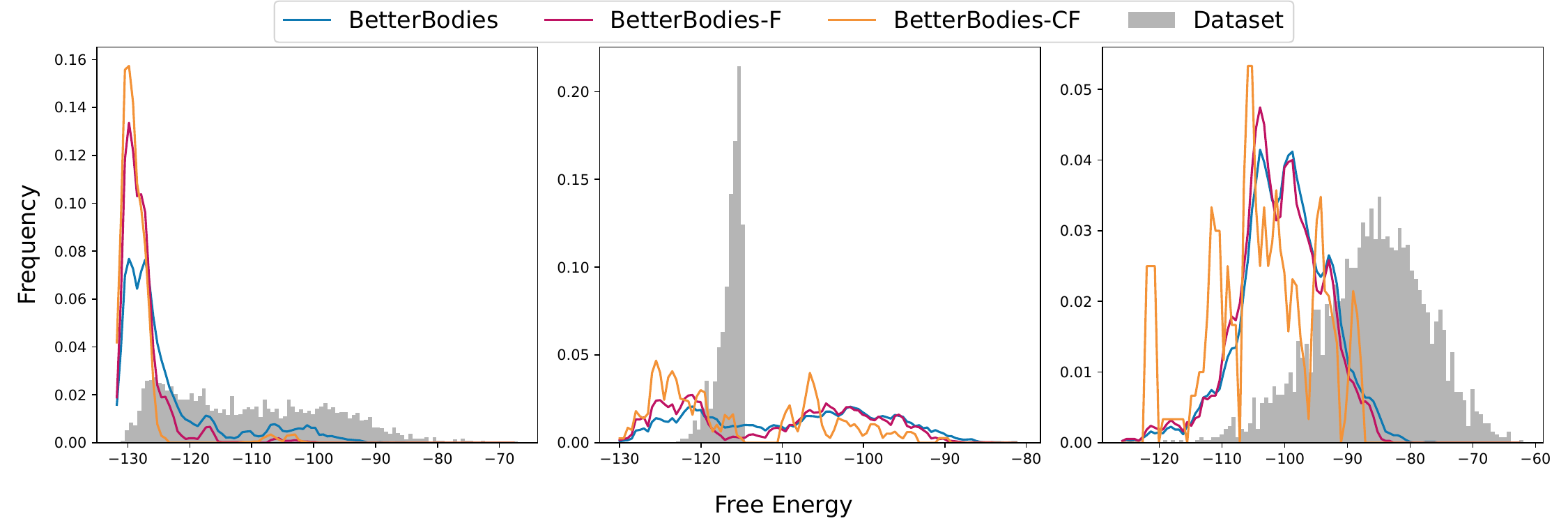}
\end{center}

\caption{Free energy distributions of unique training dataset sequences and generated sequences. The \random{} (left), \murine{} (middle), and \expert{} (right) datasets are visualized histograms. Sequences generated using BetterBodies $\eta=24$, it's F(iltering), and C(ontrastive) versions are plotted as a running average over three bins. Data is visualized as the mean over five seeds.}
\label{fig:result_dist}

\end{figure*}
Having demonstrated the efficacy of \our{} and the impact of balancing \basic{} diffusion and Q-guidance, we now present results regarding multiple diverse data distributions. We selected $\eta=24$ for \our{}, analyzing the effect of our \filtering{} and \contrastive{} method. We further compare \our{} to Basic Diffusion, where $\eta=0$, and \glspl{gflow}~\citep{jain_biological_nodate}. 

For the \filtering{} method, we include sequences above the 50th affinity percentile, scored by the Q-function. Analogously, we apply a filtering step to the sequences generated by \glspl{gflow}, including the sequences above the 50th percentile scored by the method's own learned reward model. We generate 500 novel sequences per dataset, thus returning 250 sequences after filtering.

\begin{figure*}[h]
    
\begin{center}
\begin{minipage}{0.3\textwidth}
\begin{center}
\centerline{\includegraphics[width=1\textwidth]{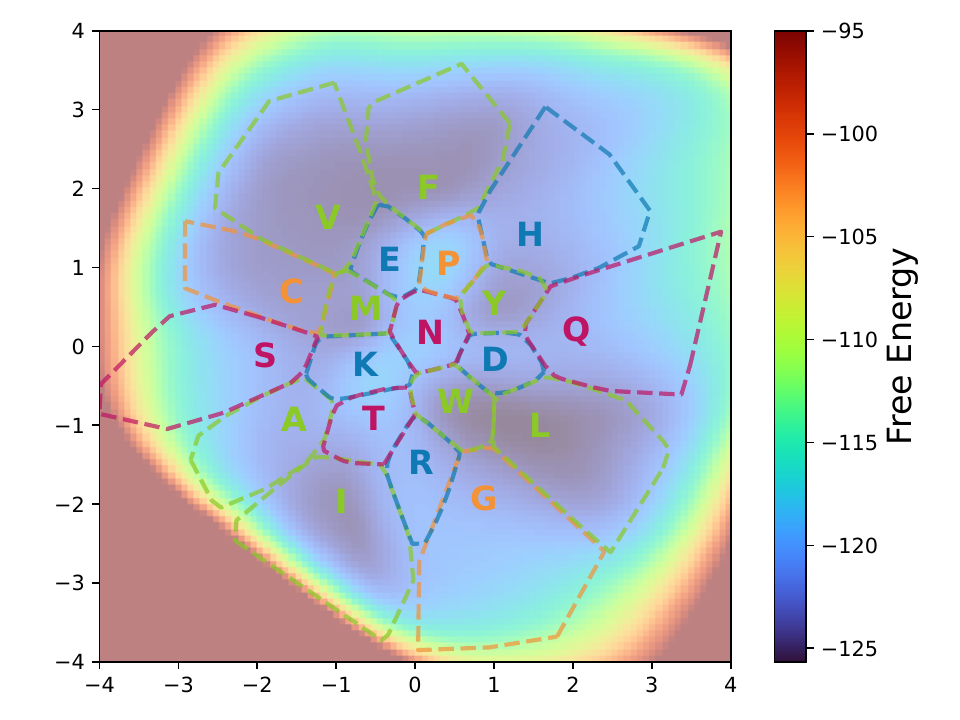}}
\end{center}
\end{minipage}
\begin{minipage}{0.3\textwidth}
\begin{center}

\centerline{\includegraphics[width=1\textwidth]{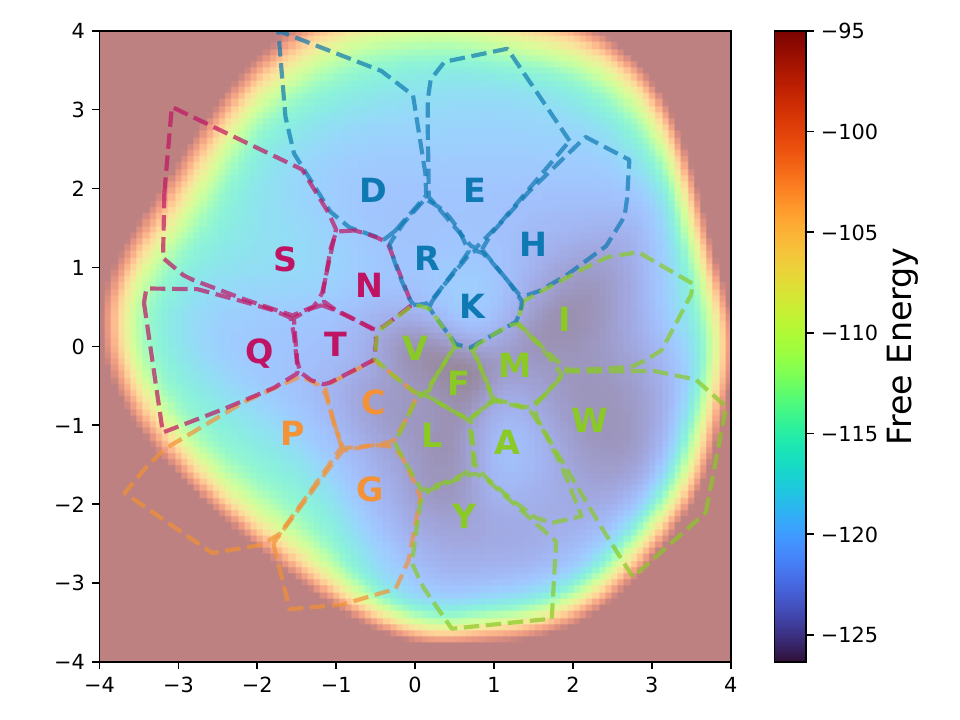}}
\end{center}
\end{minipage}
\end{center}

\caption{\gls{vae} latent space encoding amino acids, utilizing no regularization (left) and with contrastive loss regularization (right). Amino acid groups are indicated by the coloring and the space occupied by their samples. The underlying heatmap displays the average Q-value over 1000 sequence-action pairs.}
\label{fig:vae_heatmap}
\end{figure*}

In \Figref{fig:result_dist} we visualize the free energy distributions of sequences returned by \ours{} in comparison to the given dataset distribution. In \tabref{tab:Results} we give numerical results, comparing also to \glspl{gflow} and giving an insight into the novelty and diversity of generated sequences. We can observe from the distributions that \filtering{} and \contrastive{} further shift the distributions of free energies towards low free energies, indicating an improved performance. This is further supported by the Free Energy scores provided in the tabular results. In \Figref{fig:vae_heatmap} we visualize how the \contrastive{} changes the latent representations of \glspl{aa} and the corresponding average Q-values. We can observe that the \contrastive{} method allows to cluster \glspl{aa} which on average lead to better affinity scores.

We observe that \ours{} outperform or match the performance of \glspl{gflow} regarding Free Energy on all three datasets. We can further observe that novelty and diversity tend to decrease alongside decreasing free energy. Nonetheless, sequence sets generated by \glspl{gflow} exhibit a higher diversity and novelty when matching the free energy level of \ours{}.
Additionally, we observe that both \ours{} and \glspl{gflow} struggle to generate sequence sets whose mean free energy reaches that of the \murine{} dataset, due to a large fraction of low-affinity binders (cp. \Figref{fig:result_dist}(middle)). We hypothesize that this is due to the narrow distribution and lack of low-affinity samples in the dataset.
Interestingly, \glspl{gflow} which samples actions proportional to their reward in the dataset struggles with the expert dataset, while Q-learning performs well, indicating an advantage in such settings. This coincides with findings by~\citet{pmlr-v202-shen23a}, showing that low rewards were oversampled and \glspl{gflow} failed to increase the expected reward despite scaled training rewards.

\begin{table}[]
\centering

\begin{tabular}{c|cccc}
&Method&Expert&Natural&Random\\

\hline 
\multirow{6}{*}{\rotatebox[origin=c]{90}{Free Energy}} & Dataset&-110.53 $\pm$ 12.84&-116.46 $\pm$ 1.49&-86.21 $\pm$ 8.75\\
& Basic Diffusion&-105.02 $\pm$ 2.19&-109.28 $\pm$ 1.29&-84.74 $\pm$ 0.83\\
&BetterBodies&-123.23 $\pm$ 2.45&-108.40 $\pm$ 2.44&-99.64 $\pm$ 2.64\\
&BetterBodies-F&-127.44 $\pm$ 1.68&-110.53 $\pm$ 2.50&-100.55 $\pm$ 2.89\\
&BetterBodies-CF &\textbf{-128.20 $\pm$ 0.30}&\textbf{-113.40 $\pm$ 1.57}&\textbf{-103.56 $\pm$ 3.29}\\
&\glspl{gflow} & -103.85 $\pm$ 0.55&-108.11 $\pm$ 0.37&-101.28 $\pm$ 0.47\\
&\glspl{gflow}-F&-101.71 $\pm$ 0.68&-108.98 $\pm$ 0.29&\textbf{-104.02 $\pm$ 0.43}\\

\hline 
\multirow{6}{*}{\rotatebox[origin=c]{90}{Diversity}} &Dataset&7.72&7.38&10.27\\
& Basic Diffusion&7.95 $\pm$ 0.25&8.00 $\pm$ 0.22&10.17 $\pm$ 0.01\\
&BetterBodies&4.62 $\pm$ 0.33&6.16 $\pm$ 0.70&5.06 $\pm$ 0.81\\
&BetterBodies-F&4.23 $\pm$ 0.35&5.60 $\pm$ 0.73&4.55 $\pm$ 0.79\\
&BetterBodies-CF&4.22 $\pm$ 0.22&5.36 $\pm$ 0.70&6.38 $\pm$ 0.75\\
&\glspl{gflow}&9.20 $\pm$ 0.13&6.08 $\pm$ 0.09&9.24 $\pm$ 0.08\\
&\glspl{gflow}-F&9.14 $\pm$ 0.07&5.60 $\pm$ 0.13&8.77 $\pm$ 0.09\\

\hline 
\multirow{5}{*}{\rotatebox[origin=c]{90}{Novelty}} &
Basic Diffusion & 3.38 $\pm$ 0.27&2.68 $\pm$ 0.18&6.37 $\pm$ 0.04\\
&BetterBodies&2.02 $\pm$ 0.42&2.89 $\pm$ 0.44&6.24 $\pm$ 0.86\\
&BetterBodies-F&1.82 $\pm$ 0.40&2.66 $\pm$ 0.40&6.10 $\pm$ 1.15\\
&BetterBodies-CF&1.50 $\pm$ 0.07&2.59 $\pm$ 0.45&6.01 $\pm$ 0.78\\
&\glspl{gflow} &6.30 $\pm$ 0.07&5.99 $\pm$ 0.06&6.62 $\pm$ 0.02\\
&\glspl{gflow}-F &6.29 $\pm$ 0.05&5.95 $\pm$ 0.06&6.63 $\pm$ 0.04\\

\hline
\end{tabular}
    \caption{Free energy, diversity, and novelty of sequences generated by \our{}, $\eta=24$, the \filtering{} and \contrastive{} method in comparison with Basic Diffusion and \glspl{gflow}, on the \expert{}, \murine{}, and \random{} dataset. Best performing free energy values are written in bold.}
    \label{tab:Results}
\end{table}

\section{Conclusion}
We presented BetterBodies a novel method for antibody CDRH3 sequence design, demonstrating the applicability of guided latent diffusion for successive \gls{aa} sequence design. 
\Our{} successfully generates novel, diverse, and high-affinity sequences towards the SARS-CoV spike receptor-binding domain given three different sequence and affinity distributions, evaluated using the \absolut software. We demonstrated that Q-value guidance and our novel \filtering{} and \contrastive{} methods enhance the affinity of generated sequences when compared to \basic{} diffusion. 
We further demonstrate that \ours{} match or exceed the affinity scores of \glspl{gflow}, but sometimes generates less diverse sequence sets.
In conclusion, methods such as ours have the potential to have great implications for real-world biological sequence design, where the generation of novel high-affinity binders is a cost-intensive endeavor~\citep{norman2019computational, shanehsazzadeh_unlocking_2023}.

\section{Limitations and future Work}
In this work, we proposed a novel method for protein sequence generation using diffusion models and \gls{rl}. One of the main drawbacks of diffusion models is the relatively high computational time, especially for higher $N$. This could presumably be reduced using methods by~\citet{kangEfficientDiffusionPolicies2023},~\citet{nichol_improved_2021}, or~\citet{songDenoisingDiffusionImplicit2022}, which would increase the training and inference speed of \our{}.
Additionally, there are many recent methods proposed in the offline \gls{rl} community~\citep{levine2020offline} which could be used instead of clipped double Q-learning~\citep{fujimoto2018addressing}.
Finally, \our{} could be extended to the model-based and active learning setting and subsequently be evaluated using the sequence tasks proposed by~\citet{jain_biological_nodate} and~\citet{trabucco_design-bench_2022}.
\subsubsection*{Acknowledgments}
This project was funded by the Mertelsmann Foundation. This work is part of BrainLinks-BrainTools which is funded by the Federal Ministry of Economics, Science and Arts of Baden-Württemberg within the sustainability program for projects of the excellence initiative II. The authors thank Prof. Dr. Dr. h.c. mult. Roland Mertelsmann for the fruitful discussions.

\newpage
\bibliography{BetterBodies.bib}
\bibliographystyle{betterbodies}

\newpage
\appendix
\section{Appendix}
\subsection{Results on a second Antigen}
We carried out additional experiments designing antibody CDRH3 sequences binding the human CD38 (PDB ID: 3RAJ\_A), also known as cyclic ADP ribose hydrolase.
For simplicity, we reference the antigen by its PDB ID.
The datasets were retrieved as for the experiments on 2DD8, leading to datasets of size 2500, 5463, and 2103 respectively.
In \tabref{tab:3raj} we present the corresponding results.

\begin{table}[]
\centering
\begin{tabular}{c|cccc}
&Method&Expert&Natural&Random\\
\hline 

\multirow{6}{*}{\rotatebox[origin=c]{90}{Free Energy}} & Dataset&-98.39 $\pm$ 12.13&-106.09 $\pm$ 1.51&-81.90 $\pm$ 8.06\\
& Basic Diffusion &-93.66 $\pm$ 1.61&-97.77 $\pm$ 0.81&-80.69 $\pm$ 0.98\\
&BetterBodies&-107.98 $\pm$ 5.61&-103.59 $\pm$ 2.91&-92.97 $\pm$ 2.94\\
&BetterBodies-F&\textbf{-113.48 $\pm$ 6.35}&-107.13 $\pm$ 3.31&-94.04 $\pm$ 3.58\\
&BetterBodies-CF &-110.33 $\pm$ 8.87&\textbf{-110.39 $\pm$ 1.03}&-94.36 $\pm$ 2.96\\
&\glspl{gflow} &-93.46 $\pm$ 2.62&-101.21 $\pm$ 0.66&-94.58 $\pm$ 0.37\\
&\glspl{gflow} Filtered &-94.79 $\pm$ 3.21&-104.77 $\pm$ 0.89&\textbf{-96.27 $\pm$ 0.22}\\

\hline 
\multirow{6}{*}{\rotatebox[origin=c]{90}{Diversity}} & Dataset&8.30&8.06&10.27\\
& Basic Diffusion & 8.27 $\pm$ 0.12&8.23 $\pm$ 0.21&10.17 $\pm$ 0.01\\
&BetterBodies&4.91 $\pm$ 0.78&5.37 $\pm$ 0.36&6.07 $\pm$ 0.55\\
&BetterBodies-F&4.53 $\pm$ 0.66&4.64 $\pm$ 0.46&5.48 $\pm$ 0.66\\
&BetterBodies-CF &4.21 $\pm$ 0.33&4.36 $\pm$ 0.40&5.46 $\pm$ 0.70\\
&\glspl{gflow} &8.33 $\pm$ 0.19&4.69 $\pm$ 0.18&9.29 $\pm$ 0.07\\
&\glspl{gflow}-F &8.08 $\pm$ 0.22&4.28 $\pm$ 0.24&8.90 $\pm$ 0.09\\

\hline 
\multirow{5}{*}{\rotatebox[origin=c]{90}{Novelty}}
& Basic Diffusion & 3.61 $\pm$ 0.14&2.80 $\pm$ 0.22&6.37 $\pm$ 0.04\\
& BetterBodies&2.38 $\pm$ 0.63&2.82 $\pm$ 0.29&6.32 $\pm$ 0.41\\
&BetterBodies-F&2.18 $\pm$ 0.68&2.50 $\pm$ 0.24&6.18 $\pm$ 0.61\\
&BetterBodies-CF &2.66 $\pm$ 1.71&2.43 $\pm$ 0.53&5.07 $\pm$ 1.12\\
&\glspl{gflow} &5.76 $\pm$ 0.10&4.55 $\pm$ 0.10&6.63 $\pm$ 0.01\\
&\glspl{gflow}-F &5.64 $\pm$ 0.10&4.50 $\pm$ 0.10&6.63 $\pm$ 0.03\\

\hline
\end{tabular}
\caption{Antigen 3RAJ\_A; Free energy, diversity, and novelty of sequences generated by \our{}, $\eta=24$, the \filtering{} and \contrastive{} method in comparison with Basic Diffusion and \glspl{gflow}, on the \expert{}, \murine{}, and \random{} dataset. Best performing free energy values are written in bold.}
\label{tab:3raj}
\end{table}
\label{app:3raj}

\subsection{Implementation Details}
\label{app:implementation}
To reduce the effect of the latent space' structure on the reported results, we share the pre-trained \gls{vae} between all datasets for a given seed. 
Due to the large computational burden, we chose $N=5$ diffusion steps for our experiments, even though we found that $N=50$ leads to better results for $\eta=0$. This finding is analogous to~\citep{wang_diffusion_2022}.
We follow~\citep{wang_diffusion_2022} for the choice of $\beta$ noise schedule to train our diffusion model.

As in the implementation by~\citet{wang_diffusion_2022} we generate 50 actions using the Diffusion Model per step and sample the final action via a softmax distribution over the respective Q-weights.

Note, that we choose to represent $s$ for the Policy $\pi$ and Q-function as a concatenation of one-hot encodings, which represent the previous \glspl{aa}, due to its simplicity. In theory, a concatenation of \gls{vae} latent vectors, or a latent vector representing the entire sequence, could also be used.

\subsection{Amino Acid Groups}
\label{app:groups}
Our grouping of \glspl{aa} is mostly based on work by \citet{garrett_biochemistry_2010} with the following modifications:
\begin{itemize}
    \item We add the "Special Case" group
    \item we classify "P" as a special case as it ``is not an amino acid but rather an $\alpha$-imino acid.''\cite{garrett_biochemistry_2010} and `its unusual cyclic structure''\cite{garrett_biochemistry_2010}.
    \item we classify "G" as a special case as it does not have a side chain.
    \item we classify "C" as a special case as it can ``deprotonate at pH values greater than 7''\cite{garrett_biochemistry_2010}.
    \item we classify "Y" as hydrophobic as \citet{garrett_biochemistry_2010} argue that it could also be classified as such.
\end{itemize}

Note, that we chose this specific grouping not because we are convinced it bears an advantage, but rather because it was the first grouping we found in literature.
\end{document}